# Average fold-change of genetic pathways in biological transitions

Sandra Costa, Joan Nieves, Augusto Gonzalez

**Abstract**
A biological transition from a state N to a state T is characterized by a rearrangement of the gene expression profile in the system, quantitatively measured through the differential expression of genes. In contrast, changes in genetic pathways are usually evaluated by means of hypothesis testing schemes. We introduce a quantitative measure in order to evaluate the average fold-change of genetic pathways in biological transitions and applied it to characterize in general grounds the transition from a normal tissue to a tumor. Additionally, we study the transformation from primary to metastatic melanoma. Gene expression data from the TCGA portal for 16 tumors and the Reactome compilation of pathways are used for this purpose.

## 1. Introduction

Medicine has reached the molecular level. Presently, we look for genes as the ultimate causes of most diseases. Diseases are understood in terms of malfunctions of the gene regulatory network governing cell dynamics **[1]**.

Particularly in cancer, one of the hardest present challenges to medicine, genetic and epigenetic factors are known to play a significant role **[2]**. Changes in gene expression measurements in tumors are apparent when compared against the same measurements in normal samples **[3,4]**.

The profile of under- and over-expressed genes in a tumor is, however, strongly dependent on the tissue of origin. We found, for example, only 6 common genes among the first 2500 most differentially expressed genes in 15 tumors **[4]**.

Thus, the main genes are different, but are also different the processes in which they are involved? This question may be addressed by using pathway analysis **[5]**. The common procedure is the following. First, select a large enough set of the most differentially expressed genes. Second, define a test for the representation of a given genetic process or pathway **[6]** in the set of selected genes. The null hypothesis is the random distribution, that is the fraction of genes in the set corresponding to the given pathway should be proportional to the number of genes in the pathway. If the test shows probabilities much higher than predicted by chance, we say the given pathway to be enriched. Elaborations may be introduced to this procedure, but grosso modo this is the idea.

Difficulties with enrichment analysis are twofold. First, there are different compilations of genetic pathways and genes involved in them **[7]**. Still more important, the number of annotated genes is much lower than the total number of known genes. For example, there are only around 10,000 annotated genes in the Reactome database, but the number of human genes in the Ensembl compilation is 6 times larger **[8]**. The functions of a huge number of genes are unknown. This difficulty is related to the present knowledge of genetics and can be hardly overcome. The second difficulty with enrichment analysis is related to the fact that if the selected set of genes is relatively small, the result may be dependent on the used set.

In the present paper, we show a method to uniquely define the main genetic pathways involved in a biological transition. It removes the uncertainty related to the selected set of genes by using the complete gene expression profile. The idea is extremely simple: we define the average fold-change of genes in each pathway and compare it with the average fold change of all genes in the biological transition. We illustrate the method by computing the enrichment of the top (i.e. global, most generic) 28 pathways in the Reactome database for the transition from a normal tissue to the corresponding tumor. To this end, gene expression data for 15 tumors are borrowed from The Cancer Genome Atlas (TCGA) portal **[9]**.

Of course, our results in this area are not completely new. Extensive studies on gene expression data and pathway enrichment in tumors are available **[10]**. In general, the enriched pathways are related to the cancer landmarks **[11,12]**. What is remarkable in our methodology is its simplicity, the fact that it quantifies the deregulation, and the possibility to visualize similarities between groups of tumors according to the common deregulated pathways. As an example of the results, we mention that the most deregulated top pathway in the studied set of 15 tumors is related to the re-organization of the extracellular matrix, needed for the tumor invasion to its neighborhood.

The proposed methodology is applied also to the transition from primary to metastatic melanoma. It could be tested in other biological transitions where gene expression rearrangements are not so radical as in tumors. For example, the transition to Alzheimer disease in the elderly **[13]**.

**2. Average log-fold change of top genetic pathways in tumors**
As mentioned above, we use the top pathways of the Reactome database in order to perform the analysis. There are 28 such global pathways. The number of genes participating in them are shown in the diagonal of **Table I**. The off-diagonal elements refer to number of genes shared between different pathways. Notice that there are very large pathways, like Signal transduction, with thousands of genes, and small pathways, like Digestion and absorption, with only 29 genes participating in it. Notice also that there are only 10,785 annotated genes, out of the 60,483 human genes in the Ensembl compilation.

The gene expression data is borrowed from the TCGA. The 15 tumors to be analized are listed in **Table II**. In general, the number of normal samples is much less than the number of tumor samples in the data.

Normal samples are used to define reference values for the expression of each gene in a tissue, $e\_ref$. The median or the geometric average in the set of normal samples may be used for this purpose **[4]**. The log-fold variation for each gene in a given tumor sample is defined as:

$$e\_fold = Log\_2\,[e\,/\,e\_ref]. \qquad (1)$$

A value $e\_fold > 0$ means that the gene is over-expressed, whereas $e\_fold < 0$ refers to an under-expressed or silenced gene in the tissue. To measure the magnitude of deregulation, we may use the module of $e\_fold$.

Consequently, we may define the average log-fold variation of genes in a tumor sample, and in the tumor state:

$$< |e\_fold| >\_sample = \sum\_e |e\_fold| \,/\, N\_genes, \qquad (2)$$

$$< |e\_fold| >\_tumor = \sum\_samples < |e\_fold| >\_sample \,/\, N\_tumor, \qquad (3)$$

In the first case, the sum runs over genes, $N\_genes = 60483$ is the total number of measured genes, and in the second case it runs over samples in the tumor state. $N\_tumor$ is the number of such samples.

The average log-fold change of genes in the studied tumors is given in the last column of **Table II**.

In the same way, we define the average log-fold change of genes in pathways in a given tissue. First, we take sample averages, **Eq. (2)**, where the sum includes only genes participating in the pathway. Then, take the average over tumor samples, **Eq. (3)**.

The results are shown in **Table III**. The most consistently deregulated pathway in all tumors is Extracellular matrix organization, related to the invasion landmark of cancer. The next in importance seems to be DNA replication. However, there are big differences between tumors with regard to the average deregulation of this pathway, a fact that confer to it the ability to discriminate between tumors and define groups of them.

The values shown in **Table III** allow us to define groups of tumors according to the similarities in the degrees of deregulation of pathways. The results of a principal component analysis **[14]** are given in **Fig. 1**.

Three groups of tumors are apparent in this Fig. The first is conformed by PRAD and THCA, the less deregulated tumors. They are characterized by being closer to their corresponding normal states. Indeed, in gene expression space we may define a distance between the normal state and the tumor attractor **[13, 15]**. For these two tumors, the distances are the shortest in the group. A second group contains most tumors and exhibits strong deregulation of the Cell cycle, Cell-cell communication and DNA replication pathways, besides the Extracellular matrix organization, of course. Finally, a third group, comprised by READ, COAD, KIRP and KIRC shows a strong deregulation of Digestion and absorption in detriment of DNA replication and other pathways.

For illustration, we draw in **Fig. 2** the degree of deregulation of pathways in breast cancer (BRCA). In dashed red line, the average log-fold change of all genes in the tumor state is shown for reference. There are pathways, like Digestion and absorption or Metabolism of RNA, which deregulation is at the reference level. But pathways like Extracellular matrix organization or Cell cycle are well beyond the reference. They are both related to cancer landmarks.

The pathway denoted Muscle contraction is also strongly deregulated, but not related to a cancer landmark. The reason is that the tumor should also silence ordinary cellular functions in the normal state. It is apparent also in the main down-regulated genes **[4]**.

### 3. A detailed analysis of pathway enrichment in the transition from primary to metastatic melanoma

Data for skin melanoma is also borrowed from the TCGA platform. We did not include it in the analysis above because there is only 1 available normal sample. Thus, the transition from the normal state to the tumor can not be studied with a high confidence level. In addition, there are 103 primary tumor and 367 metastatic samples. We shall focus on the transition from the primary tumor to the metastatic one. Two recent references studied the same transition **[16, 17]**.

We show in **Fig. 3** the log-fold change of genes in both transitions: normal to primary tumor, and primary to metastatic. The first one can only be analyzed at the qualitative level because of the scarcity of normal samples. The x-axis of **Fig. 3** is the absolute value of the log-fold variation. The y-axis, on the other hand, is the number of genes with log-fold variation greater than or equal to a given value. We notice that there are $10^4$ genes with log-fold variation greater than one in the normal tissue to tumor transit, but only $10^3$ genes with the same characteristic in the metastasis transformation. Strongly deregulated genes are 1/6 and 1/60 of the genoma, respectively. It means that the latter is a quasi-continuous or weakly discontinuous transition **[14]**, that is a slight distortion of the tumor condition. The average log-fold variation of genes is 0.411 and 0.133, respectively.

The top deregulated pathways are depicted in **Fig. 4.** Hemostasis, Immune systems, Cell-cell communication, Extracellular matrix organization and Vesicle-mediated transport are the most significant ones, followed by Developmental Biology and Programmed cell death. All of these pathways are related to cancer landmarks and tissue physiology. The role of small vesicles in the metastasis of melanoma has been widely recognized **[18]**.

We may also analyze which particular pathways, not only the top ones, are most deregulated in the metastasis transition. To this end we start from the 2233 pathways in the Reactome database and perform the same calculation. The results are shown in **Fig. 5**. Notice that the degrees of deregulation are quantified. There is no need to select a set of genes and apply hypothesis testing or any other assumption, as in the common procedures **[5]**. In Fig. 5 it is apparent that there are around 20 strongly deregulated pathways. They are listed in **Table IV**. Pathways with less than 5 genes are excluded because they may shown artificially high log-fold variations because of the small number of genes. Probably, these are not enough studied pathways.

From the 20 pathways in the Table, which deregulation have great significance to metastatic melanoma, 14 are related to the Immune system, in accord with previous analysis. In addition, second in the ranking there is a metabolic pathway, the "Biosynthesis of melanin". It has recently been shown that melanin plays a role in suppressing numerous essential metastatic processes **[19]**. Sixth in the ranking there is the pathway "Scavenging of heme from plasma". The important role of heme in tumor cells has also been pointed out recently **[20]**. Then, in place $12^{th}$ the deregulation of the pathway "Type I hemidesmosome assembly" stresses the significance of basement membrane proteins in the epithelial-mesenchymal transition (EMT) **[21]**. Next, in place $16^{th}$ in the ranking, the pathway "Apoptotic cleavage of cell adhesion proteins", stresses the role of adhesion proteins like E-cadherin and others in EMT and the control of metastatic potential of tumors **[22]**. The pathway "Binding and Uptake of Ligands by Scavenger Receptors", $18^{th}$ in the ranking, is related to cholesterol transporting, and indirectly to the EMT transition **[23]**. Finally, the deregulation of the pathway "Formation of the cornified envelope", $19^{th}$ in the ranking, is also related to the EMT transition. Genes in the Late cornified envelope group are also likely to have tumor suppressor functions **[24]**.

A direct comparison with the results of two papers **[16, 17]** on genes and relevant pathways in the metastasis of melanoma is presented elsewhere.

**4. Concluding remarks**
We present a quantitative measure for pathway deregulation in a biological transition. The index is in fact the average of the log-fold variation of genes participating in the pathway. We apply it to characterize in general grounds the transition from a normal tissue to a tumor, and also to characterize the transformation from primary to metastatic melanoma.

In order that the results of the paper could be easily checked by an interested reader, we modified the order of averages in **Eqs. (2,3)**. First, we compute mean geometric averages of genes in the tissue. These expression vectors are deposited in the GitHub repository. Then, average gene and pathway log-fold variations are computed.

The methodology provides a unique measure allowing comparison between different tumors, for example, or the identification of enriched pathways in cases where gene expression rearrangements are not so strong as in primary tumors.

**Acknowledgments**

The authors acknowledge the Cuban Agency for Nuclear Energy and Advanced Technologies (AENTA) and the Office of External Activities of the Abdus Salam Centre for Theoretical Physics (ICTP) for support.

**Authors contributions**

A.G. conceived and coordinated the work. S.C. and J.N. processed GE data. J.N. actualized the GitHub repository. All authors analyzed and interpreted the results, contributed to the manuscript and approved the final version.

**Competing interests**

The authors declare that they have no competing interests.

**Availability of data and materials**

The information about the data we used, the procedures and results are integrated in a public repository that is part of the project "Processing and Analyzing Mutations and Gene Expression Data in Different Systems":
https://github.com/DarioALeonValido/evolp

The TCGA data for the cloud centers (tissue mean geometric averages of genes) is replicated in the path ../evolp/databases_generated/tcga-hpa/. Computation of log-fold variation of genes and pathways from these cloud centers is trivial.

|     | P1  | P2  | P3  | P4  | P5  | P6  | P7  | P8  | P9  | P10 | P11 | P12 | P13 | P14 | P15 | P16 | P17 | P18 | P19 | P20 | P21 | P22 | P23 | P24 | P25 | P26 | P27 | P28 |
|-----|-----|-----|-----|-----|-----|-----|-----|-----|-----|-----|-----|-----|-----|-----|-----|-----|-----|-----|-----|-----|-----|-----|-----|-----|-----|-----|-----|-----|
| P1  | 118 | 1   | 12  | 85  | 0   | 4   | 17  | 0   | 33  | 4   | 4   | 0   | 38  | 5   | 34  | 26  | 24  | 6   | 31  | 0   | 11  | 10  | 10  | 16  | 0   | 44  | 6   | 44  |
| P2  | 0   | 140 | 1   | 1   | 2   | 0   | 45  | 0   | 21  | 2   | 0   | 21  | 6   | 25  | 53  | 6   | 11  | 0   | 1   | 2   | 5   | 0   | 5   | 0   | 0   | 53  | 0   | 13  |
| P3  | 0   | 0   | 801 | 329 | 236 | 13  | 261 | 0   | 181 | 317 | 107 | 1   | 544 | 87  | 302 | 112 | 374 | 95  | 8   | 2   | 27  | 63  | 81  | 9   | 184 | 559 | 54  | 91  |
| P4  | 0   | 0   | 0   | 585 | 122 | 11  | 212 | 0   | 182 | 129 | 80  | 1   | 400 | 54  | 317 | 113 | 305 | 90  | 31  | 4   | 52  | 43  | 71  | 22  | 79  | 428 | 58  | 88  |
| P5  | 0   | 0   | 0   | 0   | 340 | 23  | 147 | 0   | 29  | 175 | 0   | 0   | 299 | 33  | 64  | 21  | 139 | 3   | 0   | 1   | 0   | 11  | 0   | 0   | 111 | 294 | 0   | 2   |
| P6  | 0   | 0   | 0   | 0   | 0   | 99  | 74  | 0   | 34  | 6   | 6   | 1   | 65  | 2   | 23  | 85  | 36  | 8   | 4   | 0   | 4   | 47  | 8   | 8   | 0   | 64  | 7   | 6   |
| P7  | 0   | 0   | 0   | 0   | 0   | 0   | 1135| 3   | 1   | 347 | 120 | 58  | 91  | 494 | 143 | 486 | 304 | 482 | 183 | 13  | 24  | 110 | 81  | 90  | 18  | 62  | 825 | 88  | 167 |
| P8  | 0   | 0   | 0   | 0   | 0   | 0   | 0   | 29  | 2   | 0   | 0   | 0   | 0   | 0   | 0   | 4   | 1   | 0   | 0   | 0   | 0   | 0   | 0   | 0   | 0   | 2   | 2   | 0   |
| P9  | 0   | 0   | 0   | 0   | 0   | 0   | 0   | 0   | 2116| 7   | 61  | 59  | 75  | 470 | 86  | 541 | 342 | 519 | 263 | 9   | 16  | 50  | 24  | 92  | 29  | 7   | 673 | 124 | 104 |
| P10 | 0   | 0   | 0   | 0   | 0   | 0   | 0   | 0   | 0   | 393 | 32  | 0   | 308 | 5   | 40  | 13  | 156 | 27  | 4   | 0   | 1   | 1   | 11  | 8   | 126 | 232 | 5   | 15  |
| P11 | 0   | 0   | 0   | 0   | 0   | 0   | 0   | 0   | 0   | 0   | 140 | 0   | 124 | 1   | 120 | 53  | 67  | 50  | 4   | 0   | 0   | 0   | 55  | 9   | 4   | 113 | 52  | 4   |
| P12 | 0   | 0   | 0   | 0   | 0   | 0   | 0   | 0   | 0   | 0   | 0   | 418 | 20  | 74  | 77  | 40  | 57  | 1   | 0   | 4   | 4   | 0   | 5   | 2   | 0   | 137 | 5   | 15  |
| P13 | 0   | 0   | 0   | 0   | 0   | 0   | 0   | 0   | 0   | 0   | 0   | 0   | 3183| 2   | 89  | 454 | 273 | 472 | 207 | 14  | 15  | 36  | 55  | 143 | 22  | 139 | 7   | 120 | 55  |
| P14 | 0   | 0   | 0   | 0   | 0   | 0   | 0   | 0   | 0   | 0   | 0   | 0   | 0   | 738 | 397 | 173 | 146 | 7   | 3   | 34  | 86  | 31  | 10  | 4   | 18  | 417 | 78  | 167 |
| P15 | 0   | 0   | 0   | 0   | 0   | 0   | 0   | 0   | 0   | 0   | 0   | 0   | 0   | 0   | 2260| 5   | 260 | 491 | 158 | 14  | 42  | 95  | 57  | 165 | 51  | 29  | 3   | 166 | 342 |
| P16 | 0   | 0   | 0   | 0   | 0   | 0   | 0   | 0   | 0   | 0   | 0   | 0   | 0   | 0   | 0   | 2235| 2   | 441 | 185 | 10  | 31  | 119 | 57  | 62  | 33  | 2   | 482 | 179 | 62  |
| P17 | 0   | 0   | 0   | 0   | 0   | 0   | 0   | 0   | 0   | 0   | 0   | 0   | 0   | 0   | 0   | 0   | 7   | 315 | 15  | 7   | 74  | 77  | 78  | 38  | 70  | 650 | 121 | 222 |
| P18 | 0   | 0   | 0   | 0   | 0   | 0   | 0   | 0   | 0   | 0   | 0   | 0   | 0   | 0   | 0   | 0   | 0   | 741 | 4   | 0   | 2   | 6   | 58  | 8   | 0   | 159 | 50  | 11  |
| P19 | 0   | 0   | 0   | 0   | 0   | 0   | 0   | 0   | 0   | 0   | 0   | 0   | 0   | 0   | 0   | 0   | 0   | 0   | 29  | 0   | 1   | 0   | 8   | 16  | 0   | 14  | 5   | 7   |
| P20 | 0   | 0   | 0   | 0   | 0   | 0   | 0   | 0   | 0   | 0   | 0   | 0   | 0   | 0   | 0   | 0   | 0   | 0   | 0   | 216 | 27  | 4   | 2   | 0   | 0   | 51  | 25  | 2   |
| P21 | 0   | 0   | 0   | 0   | 0   | 0   | 0   | 0   | 0   | 0   | 0   | 0   | 0   | 0   | 0   | 0   | 0   | 0   | 0   | 0   | 500 | 35  | 3   | 6   | 0   | 235 | 70  | 61  |
| P22 | 0   | 0   | 0   | 0   | 0   | 0   | 0   | 0   | 0   | 0   | 0   | 0   | 0   | 0   | 0   | 0   | 0   | 0   | 0   | 0   | 0   | 337 | 9   | 8   | 0   | 80  | 4   | 82  |
| P23 | 0   | 0   | 0   | 0   | 0   | 0   | 0   | 0   | 0   | 0   | 0   | 0   | 0   | 0   | 0   | 0   | 0   | 0   | 0   | 0   | 0   | 0   | 193 | 16  | 2   | 183 | 57  | 28  |
| P24 | 0   | 0   | 0   | 0   | 0   | 0   | 0   | 0   | 0   | 0   | 0   | 0   | 0   | 0   | 0   | 0   | 0   | 0   | 0   | 0   | 0   | 0   | 0   | 261 | 0   | 22  | 16  | 16  |
| P25 | 0   | 0   | 0   | 0   | 0   | 0   | 0   | 0   | 0   | 0   | 0   | 0   | 0   | 0   | 0   | 0   | 0   | 0   | 0   | 0   | 0   | 0   | 0   | 0   | 145 | 130 | 0   | 0   |
| P26 | 0   | 0   | 0   | 0   | 0   | 0   | 0   | 0   | 0   | 0   | 0   | 0   | 0   | 0   | 0   | 0   | 0   | 0   | 0   | 0   | 0   | 0   | 0   | 0   | 0   | 334 | 271 | 235 |
| P27 | 0   | 0   | 0   | 0   | 0   | 0   | 0   | 0   | 0   | 0   | 0   | 0   | 0   | 0   | 0   | 0   | 0   | 0   | 0   | 0   | 0   | 0   | 0   | 0   | 0   | 1   | 741 | 39  |
| P28 | 0   | 0   | 0   | 0   | 0   | 0   | 0   | 0   | 0   | 0   | 0   | 0   | 0   | 0   | 0   | 0   | 0   | 0   | 0   | 0   | 0   | 0   | 0   | 0   | 0   | 0   | 0   | 781 |

**Table I.** Number of genes in the top pathways of the Reactome database. P1 - Autophagy, P2 - Cell-Cell communication, P3 - Cell cycle, P4 - Cellular responses to external stimuli, P5 - Chromatin organization, P6 - Circadian Clock, P7 - Developmental Biology, P8 – Digestion and Absorption, P9 – Disease, P10 - DNA Repair, P11 - DNA Replication, P12 - Extracellular matrix organization, P13 - Gene expression (Transcription), P14 – Hemostasis, P15 – Immune system, P16 – Metabolism, P17 – Metabolism of proteins, P18 – Metabolism of RNA, P19 – Mitophagy, P20 – Muscle contraction, P21 – Neuronal system, P22 - Organelle biogenesis and maintenance, P23 – Programmed cell death, P24 – Protein localization, P25 – Reproduction, P26 – Signal transduction, P27 – Transport of small molecules, P28 – Vesicle-mediated transport.

| TCGA notation | Cancer type | Normal samples | Tumor samples | Average Log-fold change of genes |
|---|---|---|---|---|
| BLCA | Bladder Urothelial Carcinoma | 19 | 414 | 0.313 |
| BRCA | Breast invasive carcinoma | 112 | 1096 | 0.290 |
| COAD | Colon adenocarcinoma | 41 | 473 | 0.339 |
| ESCA | Esophageal carcinoma | 11 | 160 | 0.339 |
| HNSC | Head and and neck squamous cell carcinoma | 44 | 502 | 0.270 |
| KIRC | Kidney clear cell carcinoma | 72 | 539 | 0.377 |
| KIRP | Kidney papillary cell carcinoma | 32 | 289 | 0.323 |
| LIHC | Liver hepatocellular carcinoma | 50 | 374 | 0.308 |
| LUAD | Lung adenocarcinoma | 59 | 535 | 0.334 |
| LUSC | Lung squamous cell carcinoma | 49 | 502 | 0.426 |
| PRAD | Prostate adenocarcinoma | 52 | 499 | 0.205 |
| READ | Rectum adenocarcinoma | 10 | 167 | 0.361 |
| STAD | Stomach adenocarcinoma | 32 | 375 | 0.371 |
| THCA | Thyroid carcinoma | 58 | 510 | 0.232 |
| UCEC | Uterine corpus endometrial carcinoma | 23 | 552 | 0.379 |

**Table II.** Number of samples and average log-fold change of genes in the 15 tumors studied in the paper.

| | BLCA | BRCA | COAD | ESCA | HNSC | KIRC | KIRP | LIHC | LUAD | LUSC | PRAD | READ | STAD | THCA | UCEC |
|---|---|---|---|---|---|---|---|---|---|---|---|---|---|---|---|
| P1 | 0.436 | 0.464 | 0.443 | 0.438 | 0.366 | 0.399 | 0.398 | 0.730 | 0.444 | 0.595 | 0.238 | 0.502 | 0.392 | 0.244 | 0.495 |
| P2 | 0.940 | 0.771 | 0.719 | 1.074 | 0.798 | 0.545 | 0.534 | 1.165 | 0.841 | 1.276 | 0.341 | 0.644 | 0.931 | 0.287 | 1.103 |
| P3 | 0.912 | 0.838 | 0.676 | 1.017 | 0.727 | 0.604 | 0.564 | 0.956 | 0.825 | 1.173 | 0.407 | 0.622 | 0.790 | 0.365 | 1.051 |
| P4 | 0.681 | 0.626 | 0.622 | 0.747 | 0.554 | 0.520 | 0.511 | 0.801 | 0.611 | 0.828 | 0.345 | 0.628 | 0.569 | 0.338 | 0.797 |
| P5 | 0.786 | 0.712 | 0.500 | 0.805 | 0.582 | 0.441 | 0.377 | 0.784 | 0.661 | 0.849 | 0.426 | 0.462 | 0.532 | 0.262 | 0.876 |
| P6 | 0.568 | 0.502 | 0.566 | 0.533 | 0.491 | 0.552 | 0.564 | 0.589 | 0.437 | 0.600 | 0.344 | 0.637 | 0.458 | 0.514 | 0.699 |
| P7 | 0.652 | 0.647 | 0.607 | 0.674 | 0.569 | 0.590 | 0.554 | 0.597 | 0.591 | 0.880 | 0.405 | 0.674 | 0.542 | 0.430 | 0.783 |
| P8 | 0.276 | 0.338 | 1.294 | 1.158 | 0.340 | 0.696 | 0.728 | 0.603 | 0.530 | 0.778 | 0.309 | 1.325 | 0.872 | 0.135 | 0.362 |
| P9 | 0.574 | 0.557 | 0.604 | 0.643 | 0.521 | 0.672 | 0.660 | 0.703 | 0.593 | 0.798 | 0.359 | 0.653 | 0.538 | 0.383 | 0.697 |
| P10 | 0.808 | 0.695 | 0.544 | 0.902 | 0.637 | 0.422 | 0.385 | 0.862 | 0.684 | 0.924 | 0.355 | 0.471 | 0.650 | 0.257 | 0.852 |
| P11 | 0.962 | 0.734 | 0.744 | 1.227 | 0.861 | 0.507 | 0.522 | 1.197 | 0.864 | 1.353 | 0.265 | 0.650 | 0.918 | 0.277 | 1.014 |
| P12 | 1.061 | 1.064 | 0.964 | 1.085 | 1.220 | 1.025 | 0.981 | 0.732 | 0.979 | 1.334 | 0.558 | 1.026 | 0.950 | 0.820 | 1.218 |
| P13 | 0.610 | 0.534 | 0.542 | 0.664 | 0.512 | 0.448 | 0.450 | 0.726 | 0.538 | 0.750 | 0.315 | 0.546 | 0.564 | 0.344 | 0.710 |
| P14 | 0.709 | 0.755 | 0.930 | 0.707 | 0.662 | 0.863 | 0.810 | 0.745 | 0.879 | 1.068 | 0.432 | 0.992 | 0.638 | 0.481 | 0.925 |
| P15 | 0.592 | 0.603 | 0.734 | 0.694 | 0.628 | 0.785 | 0.645 | 0.665 | 0.701 | 0.905 | 0.349 | 0.767 | 0.583 | 0.434 | 0.716 |
| P16 | 0.590 | 0.625 | 0.693 | 0.596 | 0.547 | 0.730 | 0.732 | 0.734 | 0.605 | 0.820 | 0.405 | 0.725 | 0.524 | 0.418 | 0.727 |
| P17 | 0.581 | 0.537 | 0.579 | 0.628 | 0.506 | 0.547 | 0.540 | 0.713 | 0.594 | 0.770 | 0.354 | 0.605 | 0.533 | 0.343 | 0.669 |
| P18 | 0.458 | 0.365 | 0.565 | 0.647 | 0.438 | 0.339 | 0.330 | 0.786 | 0.447 | 0.660 | 0.274 | 0.527 | 0.556 | 0.203 | 0.430 |
| P19 | 0.505 | 0.425 | 0.485 | 0.522 | 0.410 | 0.389 | 0.420 | 0.687 | 0.430 | 0.704 | 0.266 | 0.507 | 0.476 | 0.218 | 0.457 |
| P20 | 1.035 | 0.871 | 0.804 | 0.653 | 0.826 | 0.774 | 0.895 | 0.509 | 0.753 | 1.019 | 0.615 | 1.034 | 0.685 | 0.520 | 1.189 |
| P21 | 0.683 | 0.601 | 0.655 | 0.602 | 0.482 | 0.725 | 0.726 | 0.532 | 0.616 | 0.835 | 0.438 | 0.776 | 0.553 | 0.505 | 0.841 |
| P22 | 0.472 | 0.414 | 0.514 | 0.545 | 0.442 | 0.465 | 0.409 | 0.743 | 0.468 | 0.630 | 0.259 | 0.507 | 0.485 | 0.309 | 0.587 |
| P23 | 0.606 | 0.568 | 0.554 | 0.775 | 0.545 | 0.562 | 0.478 | 0.701 | 0.533 | 0.888 | 0.278 | 0.557 | 0.527 | 0.327 | 0.681 |
| P24 | 0.449 | 0.420 | 0.568 | 0.499 | 0.404 | 0.570 | 0.566 | 0.677 | 0.448 | 0.613 | 0.337 | 0.546 | 0.436 | 0.254 | 0.553 |
| P25 | 0.838 | 0.749 | 0.508 | 0.857 | 0.607 | 0.475 | 0.394 | 0.694 | 0.721 | 0.879 | 0.441 | 0.434 | 0.586 | 0.268 | 0.965 |
| P26 | 0.651 | 0.639 | 0.603 | 0.642 | 0.528 | 0.642 | 0.620 | 0.597 | 0.607 | 0.827 | 0.372 | 0.656 | 0.555 | 0.413 | 0.787 |
| P27 | 0.617 | 0.669 | 0.811 | 0.673 | 0.633 | 0.992 | 0.996 | 0.657 | 0.647 | 0.887 | 0.427 | 0.843 | 0.618 | 0.453 | 0.810 |
| P28 | 0.539 | 0.601 | 0.717 | 0.584 | 0.520 | 0.642 | 0.558 | 0.782 | 0.703 | 0.807 | 0.345 | 0.773 | 0.554 | 0.424 | 0.688 |

**Table III.** The average logarithmic fold-change of genes in the top pathways for the 15 tumors under study.

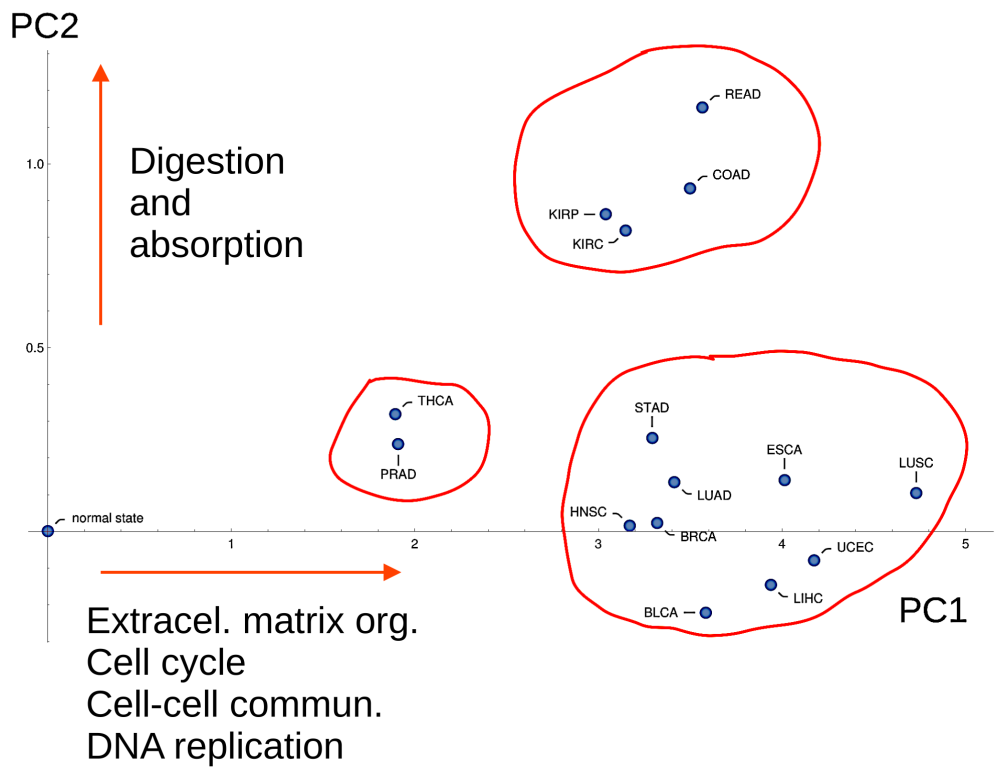

**Fig. 1.** Principal Component Analysis of the average logarithmic fold-change of pathways in tumors. Three clusters of tumors are apparent in the figure. The main deregulated pathways along each PC axis are indicated.

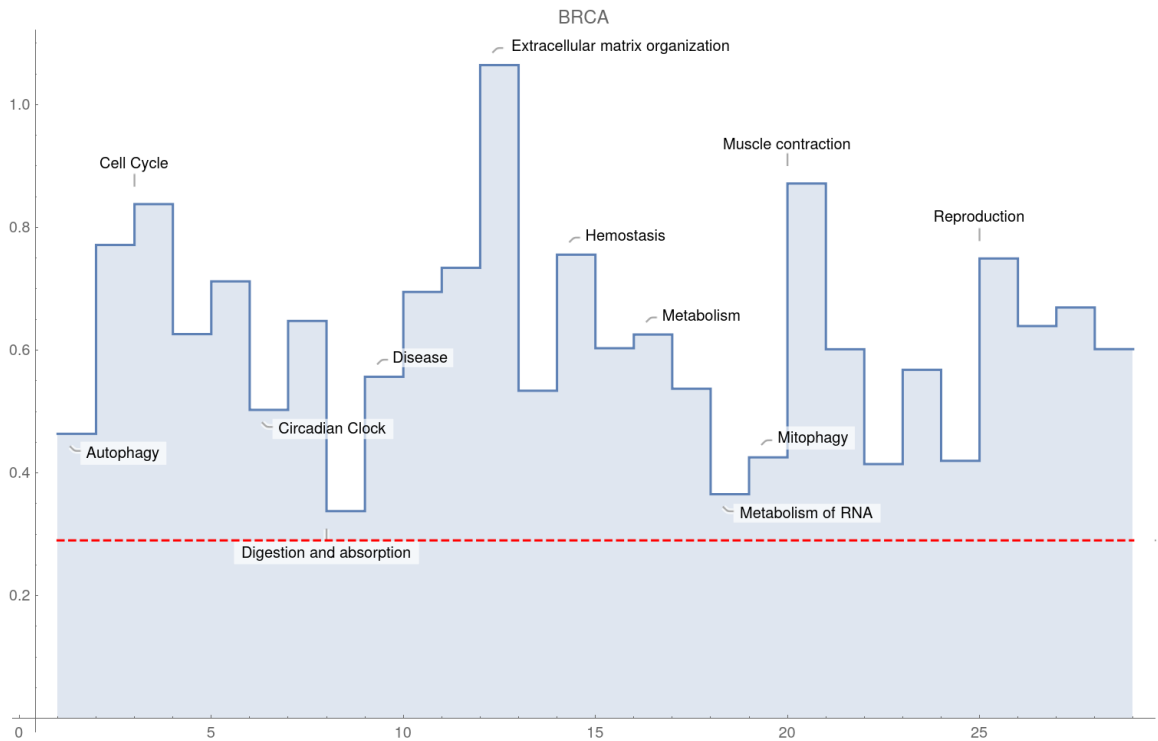

**Fig. 2.** The average logarithmic fold-change of genes in top pathways in BRCA. The mean fold-change of all genes is drawn in red for reference. Notice the 3 most deregulated pathways in this tumor: Extracellular matrix organization, Muscle contraction and Cell cycle.

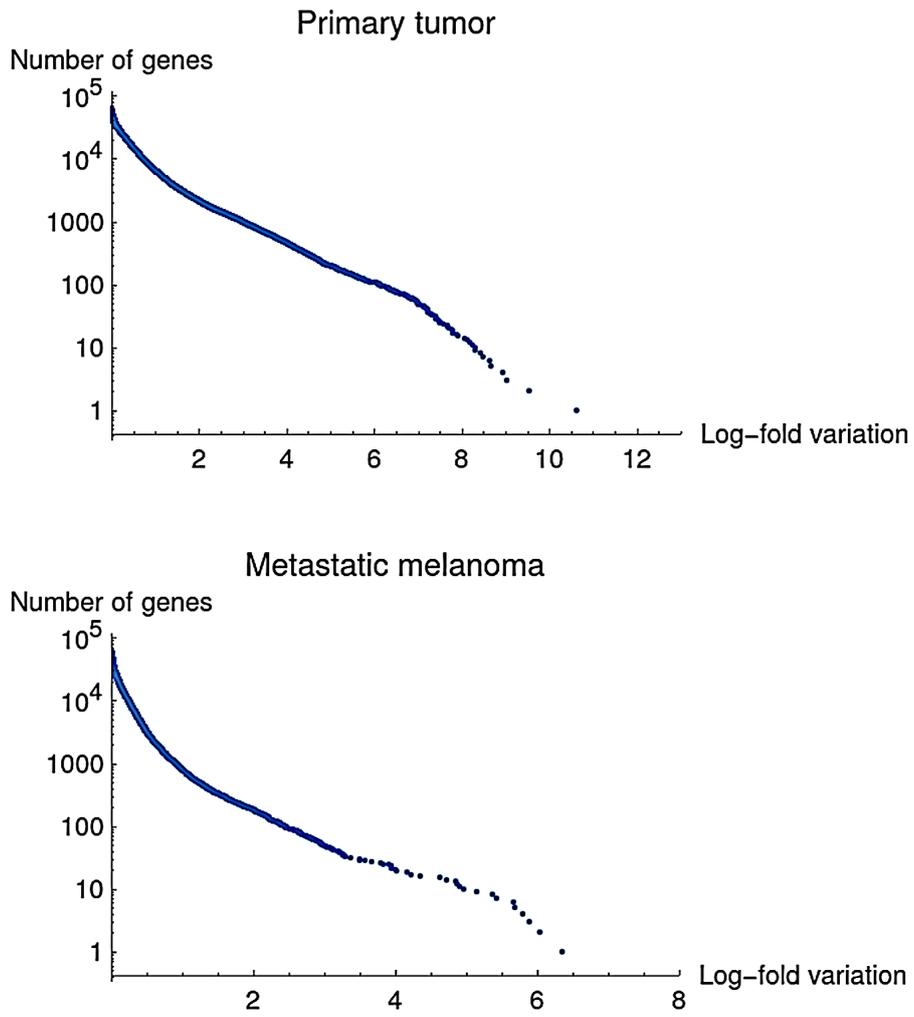

**Fig. 3**. The number of genes with absolute value of the log-fold variation greater or equal to a given value in both the primary tumor and metastatic melanoma.

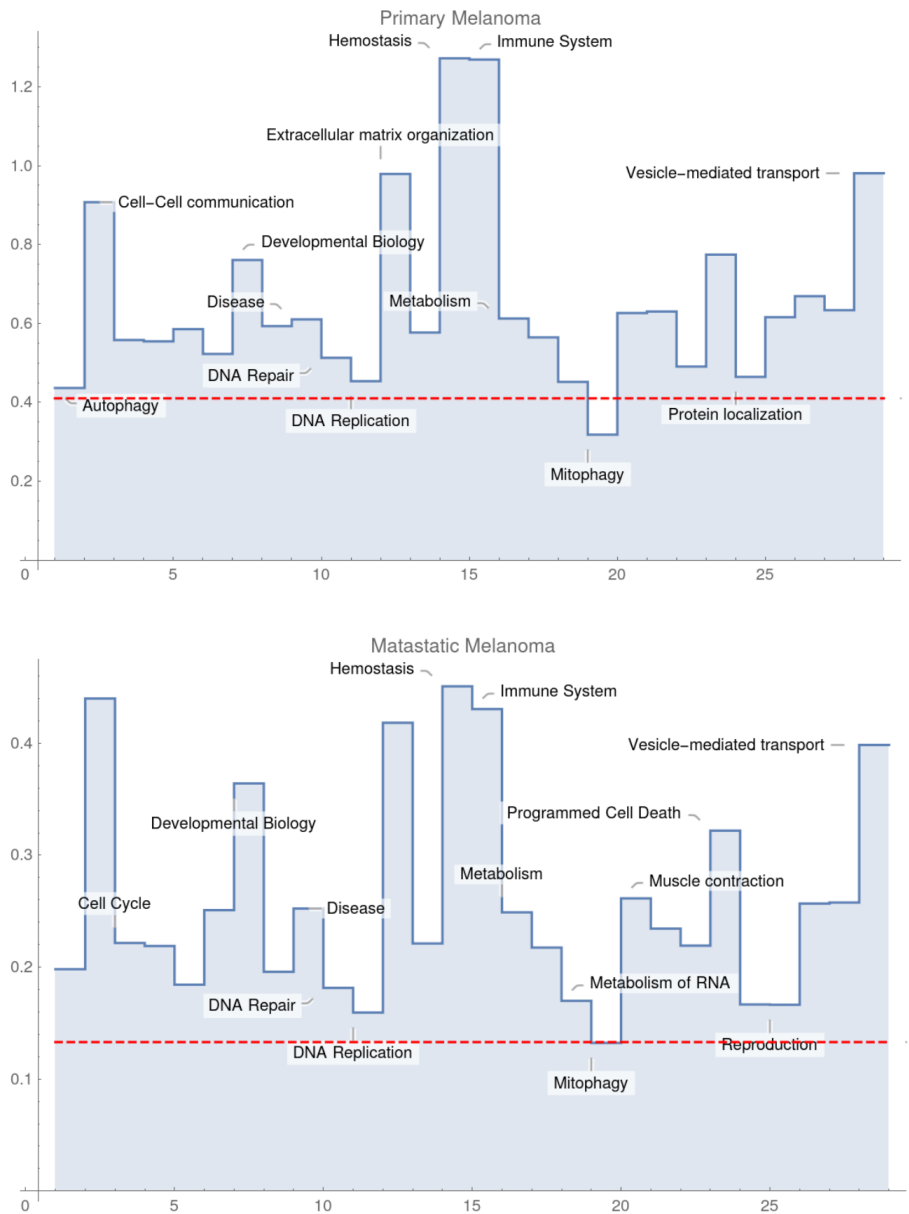

**Fig. 4.** Top pathways deregulation in the normal tissue to primary melanoma and primary to metastatic tumors transitions. The average log-fold variation of genes are drawn as red dashed lines.

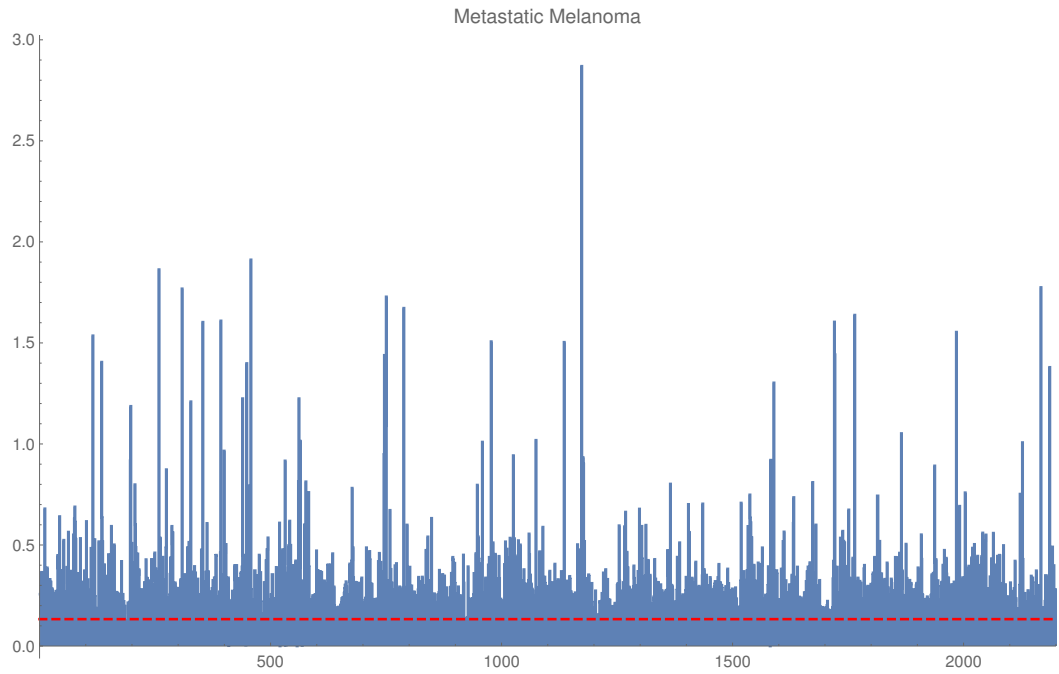

**Fig. 5.** Degrees of deregulation of the 2233 Reactome pathways in metastatic melanoma. The average log-fold change of genes is drawn as a dashed red line. The 20 most deregulated pathways are listed in **Tab. IV**.

| Rank | Expression | PathwayName | Number of genes | Top pathway |
|---|---|---|---|---|
| 1 | 3.27 | Metal sequestration by antimicrobial proteins | 6 | Immune system |
| 2 | 3.12 | Melanin biosynthesis | 5 | Metabolism |
| 3 | 2.85 | CD22 mediated BCR regulation | 61 | Immune system |
| 4 | 2.80 | Classical antibody-mediated complement activation | 63 | Immune system |
| 5 | 2.73 | FCGR activation | 132 | Immune system |
| 6 | 2.60 | Scavenging of heme from plasma | 69 | Vesicle-mediated transport |
| 7 | 2.58 | Creation of C4 and C2 activators | 71 | Immune system |
| 8 | 2.52 | Role of LAT2/NTAL/LAB on calcium mobilization | 136 | Immune system |
| 9 | 2.46 | Initial triggering of complement | 79 | Immune system |
| 10 | 2.40 | Antigen activates B Cell Receptor (BCR) leading to generation of second messengers | 163 | Immune system |
| 11 | 2.34 | Role of phospholipids in phagocytosis | 83 | Immune system |
| 12 | 2.33 | Type I hemidesmosome assembly | 11 | Cell-cell communication |
| 13 | 2.30 | FCERI mediated Ca+2 mobilization | 162 | Immune system |
| 14 | 2.13 | Regulation of Complement cascade | 103 | Immune system |
| 15 | 2.10 | FCERI mediated MAPK activation | 747 | Immune system |
| 16 | 2.04 | Apoptotic cleavage of cell adhesion proteins | 11 | Programmed cell death |
| 17 | 2.01 | Complement cascade | 114 | Immune system |
| 18 | 2.01 | Binding and Uptake of Ligands by Scavenger Receptors | 111 | Vesicle-mediated transport |
| 19 | 1.96 | Formation of the cornified envelope | 130 | Developmental biology |
| 20 | 1.84 | Fcgamma receptor (FCGR) dependent phagocytosis | 239 | Immune system |

**Table IV**. The first 20 most deregulated pathways in metastatic melanoma. Pathways with less than 5 genes are excluded from the analysis.